# Experimenting with Request Assignment Simulator (RAS)

R. Arokia Paul Rajan
Research Scholar
Department of Computer Science and Engineering
Pondicherry Engineering College, Pondicherry, India.
paulraajan@gmail.com

F. Sagayaraj Francis
Professor
Department of Computer Science and Engineering
Pondicherry Engineering College, Pondicherry, India.
fsfrancis@pec.edu

*Abstract*— There is no existence of dedicated simulators on the Internet that studies the impact of load balancing principles of the cloud architectures. Request Assignment Simulator (RAS) is a customizable, visual tool that helps to understand the request assignment to the resources based on the load balancing principles. We have designed this simulator to fit into Infrastructure as a Service (IaaS) cloud model. In this paper, we present a working manual useful for the conduct of experiment with RAS. The objective of this paper is to instill the user to understand the pertinent parameters in the cloud, their metrics, load balancing principles, and their impact on the performance.

I. INTRODUCTION

Cloud computing has changed the way in which enterprises planned for their IT infrastructure. It transformed the resources as consumable services categorized in three levels such as infrastructure, platform, and applications. Cloud computing is not a technology but, a model of delivering the services to the users. The services are consumed on demand and it can be scaled up or down. It uses pay-for-usage consumption model [1]. The resources are consumed as a web service over a network. The cloud makes the users feel that the resources are limitless, available at minimal cost, and also reliable. It relieves the user not to worry about how stuffs are working, how it is constructed, who operates it, and where it is located. When a service is consumed through the cloud, it can be ubiquitously accessed often at lower cost. It leads to the possibility of enhanced collaboration, analysis, and integration on a shared platform. Cloud wrappers the services in the internet composed with dissimilar architectural layers. Cloud architectures catering global requests for the services administered by many service broker policies.

In large scale distributed computing architectures like cloud, load balancing principle distributes the requests across a pool of computing resources. These resources may include a computer, group of computers, network links, central processing units or disk drives [2]. Load balancing principles aim to optimize resource utilization, maximizing throughput, minimizing response time, and avoid overloading of any resource. Load balancing principles use numerous scheduling algorithms to determine the assignment of servers for the requests. To name a few of these algorithms are random assignment, round robin, first-in, first-out, shortest remaining time, fixed priority preemptive scheduling, and multilevel queue scheduling [3]. Request assignment is one of the challenging task that is having a major impact on the performance of the system. Load balancing strategies actually decides this principle of assignment.

As a part of our research, we developed a novel model for experimenting the load balancing principles, namely Request Assginment Simulator (RAS). RAS is designed to fit into IaaS cloud architectures. Experimenting with RAS can be a test bed for any web service provider who is planning to subscribe IaaS. Also, this simulator will be a useful tool for the researchers and application designers who are studying the nature of service broker policies [4]. To the best of our knowledge, there is no existence of dedicated simulators to experiment cloud load balancing principles.

We developed RAS simulator using .Net framework with C# and MS Access. The user can install the application by downloading the source file from the link: http://www.paulraajan.blogspot.in. After the successful installation, the user is ready to experiment with RAS. Fig.1 shows the step-by-step method of conducting an experiment with RAS.

The rest of the paper is organized as follows: Section 2 discusses few cloud simulators with their purposes. Section 3 elaborately discusses about different components of RAS as well as how to carry out the experiment. Section 4 gives a conclusion with future expansion possibilities.

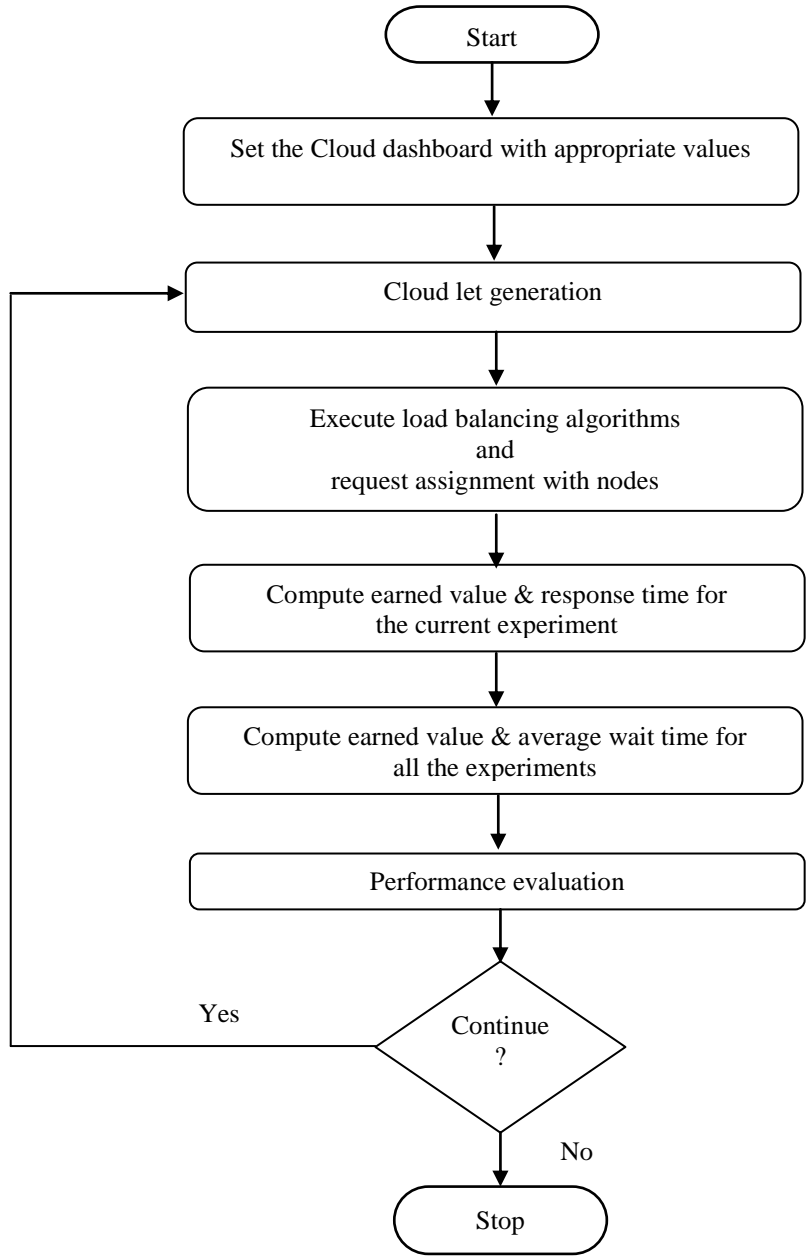

Figure 1. Flow of execution of RAS

## II.   RELATED WORKS

CloudSim is a toolkit with Java library simulates cloud computing scenarios [5]. It includes the classes for characterizing data centers, services, virtual machines, users, computing resources, and service broker policies for managing different parts of the system. This tool can also be used to evaluate the efficiency of strategies from different objective functions. It is not a ready-to-use solution where you set the parameters and collect results. Being a library, CloudSim requires that the user has to write Java programs to compose the desired scenario. CloudSim fits a better model for Infrastructure as a Service (IaaS), Platform as a Service (PaaS) and Software as a Service (SaaS) service provider's infrastructure.

Cloud Analyst is a tool that supports evaluation of social network, according to geographic distribution of data centers and users [6]. This tool simulates communities of users and data centers based on their location; parameters of a social network application and requests load in the data center. CloudAnalyst is a visual platform for the user which ease the user to configure the cloud infrastructures. CloudAnalyst is a modeling simulator built on the CloudSim framework with an extension of more capabilities. Users can set up a scenario with the goal of optimizing the system's performance by suitably opting the service broker and load balancing principles.

SimGrid [7], MicroGrid [8], GangSim [9] are grid simulators which are effective in evaluating the performance of the applications in a distributed environment. Simulation Program for Elastic Cloud Infrastructures (SPECI) is a tool that allows the prediction of the performance and behavior of data centers [10]. This tool specifically evaluates the scaling properties of large data center behavior, based on the varying inputs from middleware policies. Green Cloud is a simulation tool for energy-aware data centers for cloud computing [11]. This tool is designed to simulate workload distribution, energy consumed by data center components like servers, switches, and links as well as packet-level communication patterns. Cloud Deployment Options Simulator (CDOSim) helps in evaluating the challenges incorporated with migration of software systems to cloud environments [12]. It simulates costs of a CDO, Service Level Agreements (SLA) violations, and performance metrics, run time adaptation strategies, selection of a cloud provider, instances configuration and component deployment of virtual machines. DCSim Data Centre Simulation tool is an extensible framework for modeling dynamically managed resources in a virtualized, multi-tenant data center [13]. TeachCloud is a modeling and simulation environment for cloud computing [14]. Learners can experiment with: data centers, computing elements, networking, Service Level Agreement (SLA) constraints, and virtualization.

### III. HOW TO EXPERIMENT WITH RAS

In this section, we present the working manual for conducting an experiment with RAS. Fig. 2 shows the startup screen of the simulator.

**Start-up window:**

The main window contains the menu bar and a tool bar. From the menu bar, a user can create a new experiment or open an existing experiment which has been conducted earlier. When the user wants to open an existing file, click 'Open' option. This window lists all the existing experiments conducted so far by the users. Information about when the experiment was created also presented. The user can choose the experiment, whichever he wants to open.

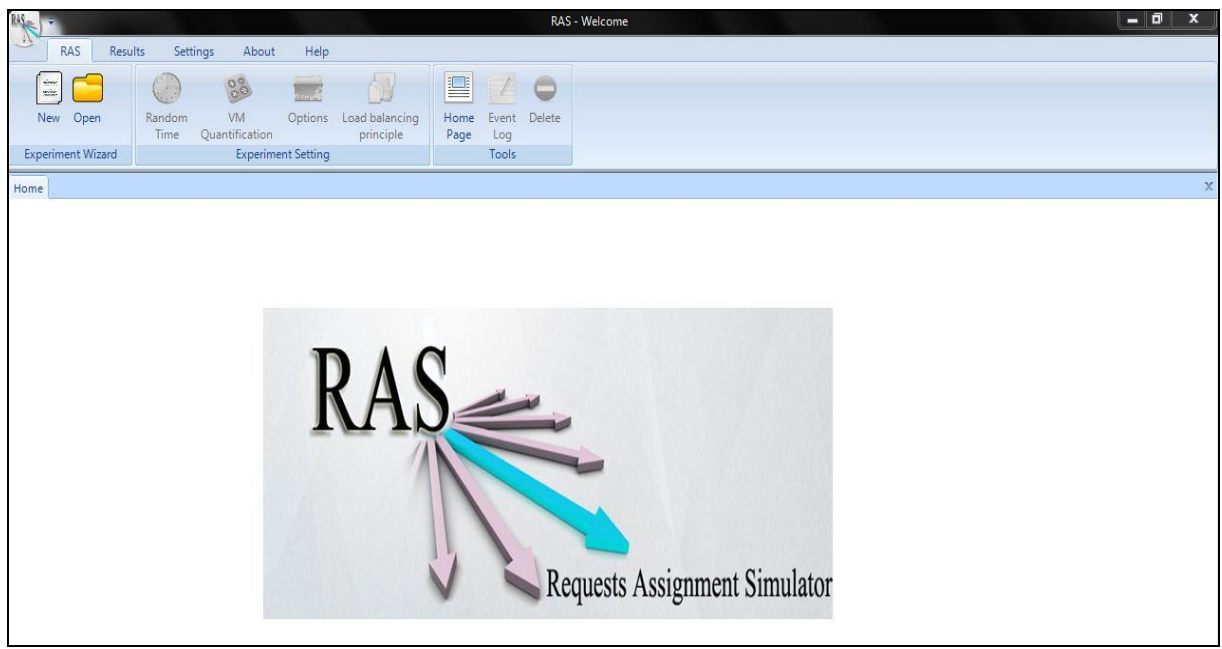

Figure 2. Start up window of RAS

In RAS, the user can create any number of experiments within a file. Each file is given with a name by the user. Experiments will be automatically assigned with unique identification numbers. This feature enables the user to set up a configuration and then repeat the experiments with varied inputs. For example, 'Test' may be a file name that contains a set of experiments. When the user creates a new file, he is supposed to configure the values for various parameters which are presented in the cloud dashboard. Fig. 3 presents the cloud dashboard that shows the master configuration setup for the parameters. These parameters are originating from the four influencing domains of cloud architectures namely, data centers, virtual machines, services, and requests. The event log will track the user's activities from the start of the current execution of an experiment.

A. *Cloud dashboard*

Cloud dashboard is the master setup window. It shows the current experiment ID of the opened file. CD contains the four important components which represent the cloud architectures. They are, 1. Data center profile 2. VM configuration 3. Services 4. Requests

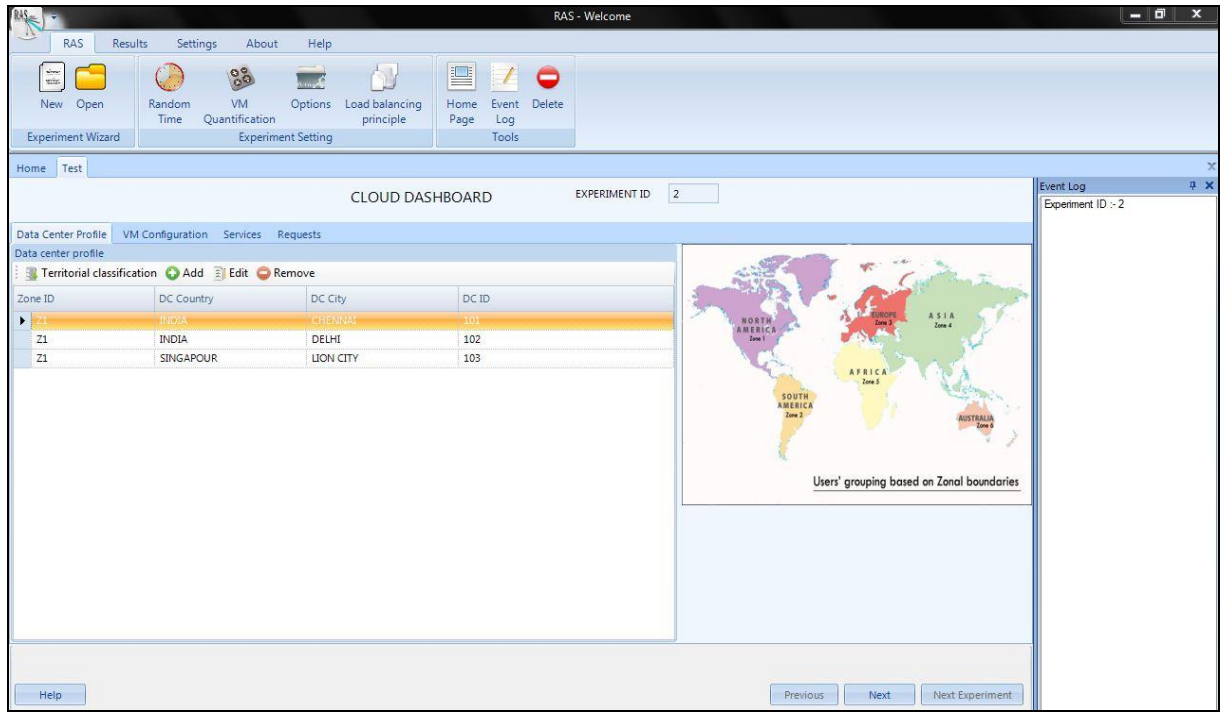

Figure 3. Cloud dashboard

a) *Data center panel*

Data center profile gives the territorial classification of the data centers. Our model assumes the World divided into six zones. The zone classification is shown in the map. Each data center is characterized by its zone ID, data center ID, country and city on which it is located. Fig. 4. presents the data center profile maintenance. If the user selects a zone ID, it automatically pops up the continent. The user can add, edit or remove a data center.

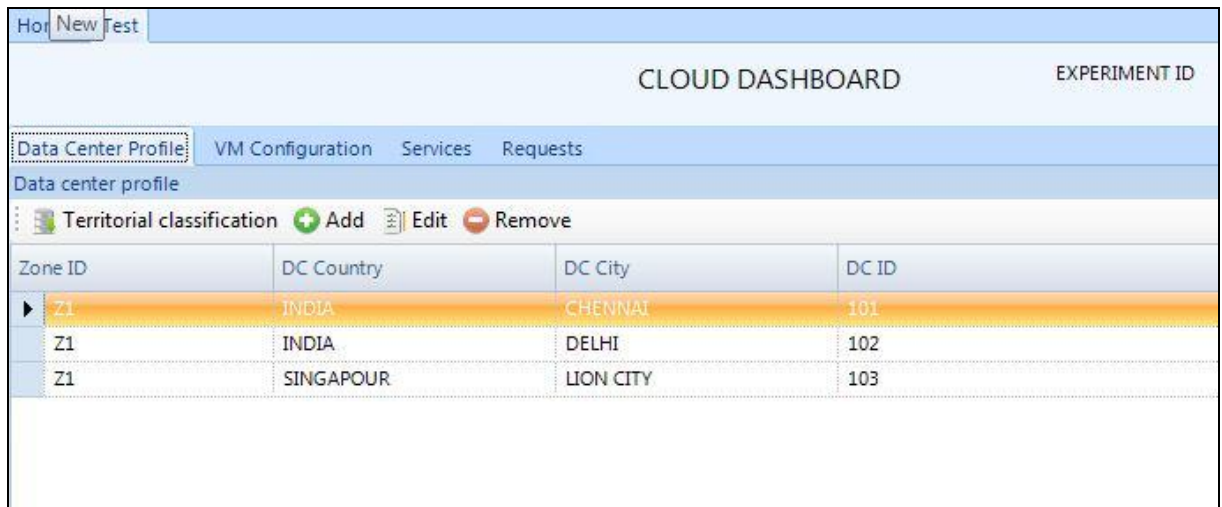

Fig. 4. Details about the data centers

b) *VM Configuration panel*

Fig. 5 shows the virtual machine configuration contains the important parameters of a virtual machine. We have taken storage capacity and load capacity as the limiting constraints for allocation of request to a virtual machine. For getting significant differences in the results, the user has to input RAM size (in GB) less than the No of connections value.

Figure 5. Configuration setup of the virtual machines

*c) Services panel*

Fig. 6 presents the window simulates the services available in the physical layer. Each service is embedded in a container file. Each service is described by its ID, name, and type. Each service is assigned with a value automatically based on the number of requests set by the user. For example, if there are three services A, B, and C with a request of 35, 23, and 42 correspondingly, then A is assigned with value 2, B with 1 and C with 3. Weightage of a service is assigned by the application designer. For example, in a reservation system, request that seeks a search service is given relatively lesser weightage comparing with the request seeking transaction service. Value and waited for the services are used for assigning priority for the requests.

Our model assumes that the services are embedded inside a container.The container may be a web page. For example, if a user wants to book a ticket for a bus, he has to go to the web page and enter the booking details. Once the details are submitted, it is sent to the server as parameters and then an instantiate an object for that request. Similarly, different objects are instantiated for different services as the requests are populated.

Figure 6. Details about the available services

*d) Requests panel*

Using this window, the user can assign consolidated requests for each service. Once the requests are given, RAS generates the pool of requests which simulates the requests originating from worldwide. RAS actually simulates the cloud let creation in a cloud by simulating random request generation for services. Fig. 7 shows the assignment of total number of requests for the services. For example, if the service ID 501 is set with fourteen requests at time *t0* is simulated to produce fourteen requests with different arrival time. After setting up the number of requests for each service at the service panel, the user has to click 'next' to proceed further.

Figure 7. Assigning requests for the services

R. Arokia Paul Rajan et al. / International Journal on Computer Science and Engineering (IJCSE)

### B. Requests Generation:

Requests panel is used to create a set of requests. RAS randomly generates the IP address of a request origin, arrival time, and process time. Fig. 8 shows such a randomized generation of inputs.

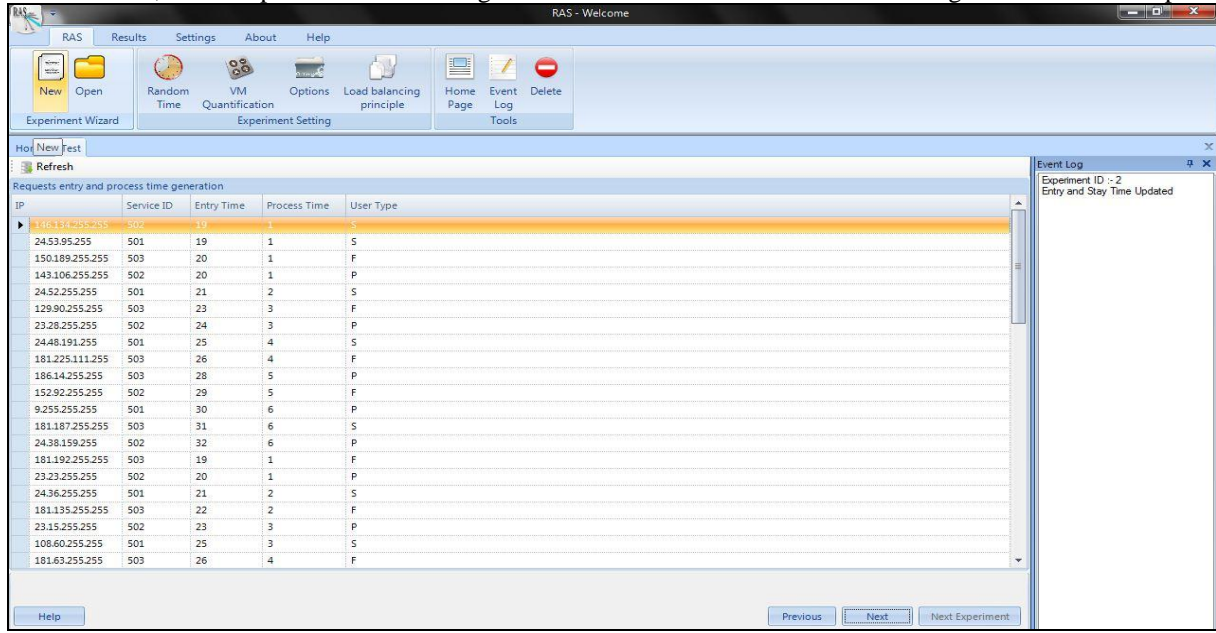

Figure 8. Generation of random inputs

If the user wants to get a new set of inputs, he can refresh the window and get the new set of inputs without changing the total number of requests. Each user is identified by the IP address. RAS actually randomly generates the IP addresses. Each IP address is having its geographical grouping with zones. For example, IP address 115.241.18.210 is originating from India. Each IP address represents a request from a user. Each request is generated with a random arrival time and process time. RAS allows the user to adjust the entry and process time settings.

### C. Execution of load balancing principles

Before generation of requests, the user is supposed to select the strategies of his preference. RAS will execute the load balancing principles selected by the user for the current experiment. If the user not opting the strategies, then the application executes round robin and orderly circular principles defaultly. Fig. 9 shows the assignment of requests with the virtual machines. Processing all the strategies for more requests may take some time. Because, RAS assigns the requests with virtual machines based on its arrival time without violating the capacity constraints.

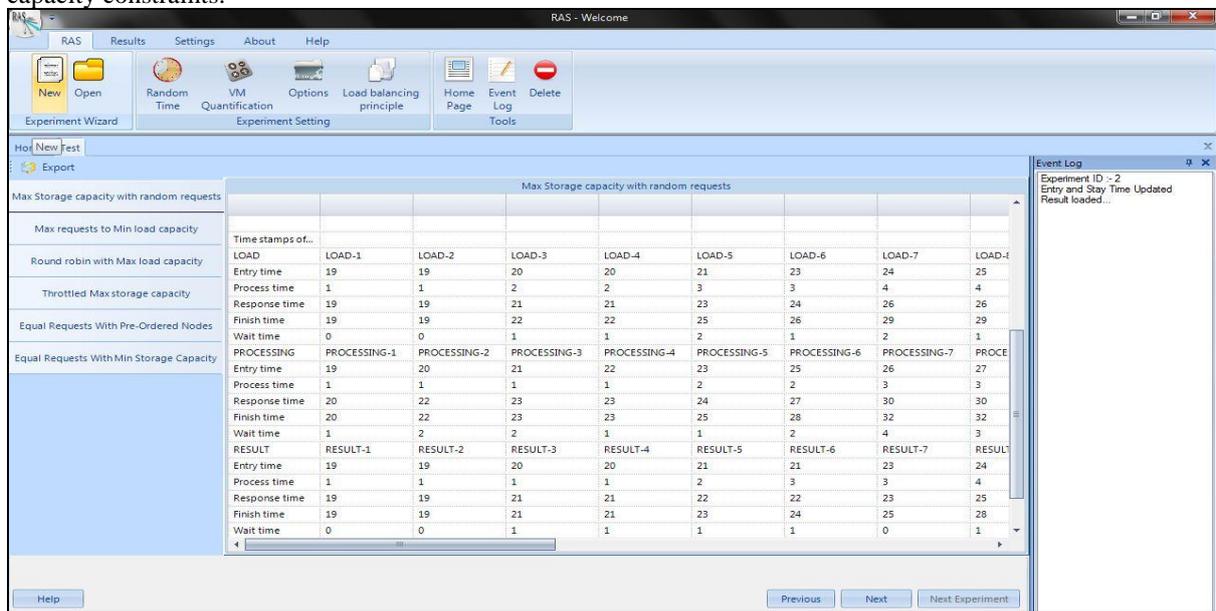

Figure 9. Assignment of requests with nodes

*D. Computation of parameters:*

After the successful execution of the chosen load balancing principles, RAS computes request's wait time, response time, total value earned by each node, and total value earned by each node against each service. This computation is carried out for the current experiment only. Fig. 10. presents the computation of various parameters.

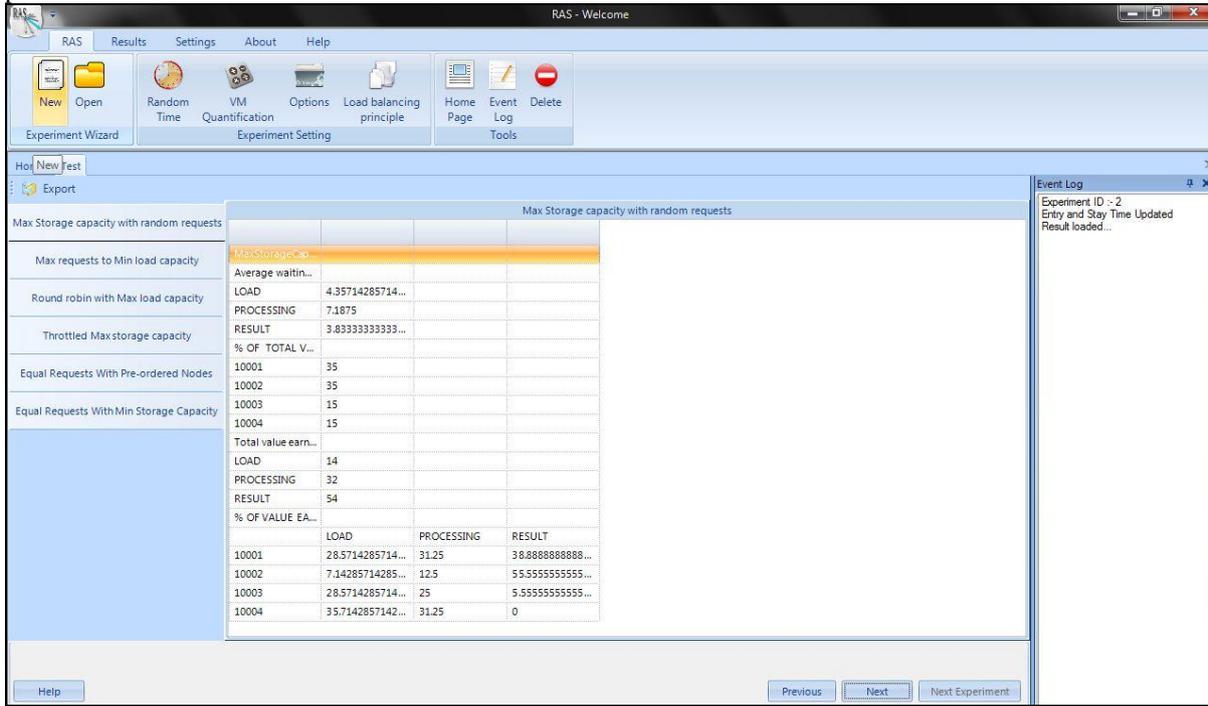

Fig. 10. Calculation of parameters

*E. Comparison of output parameters for the current experiment*

RAS produces a comparison chart for an average wait time and value earned by each virtual node. This chart plots the calculated parameters for the current experiment. Using this chart, the user can understand the performance achieved by different load balancing principles for the same set of inputs. Currently RAS supports 57 types of charts and a sample is shown in Fig. 11.

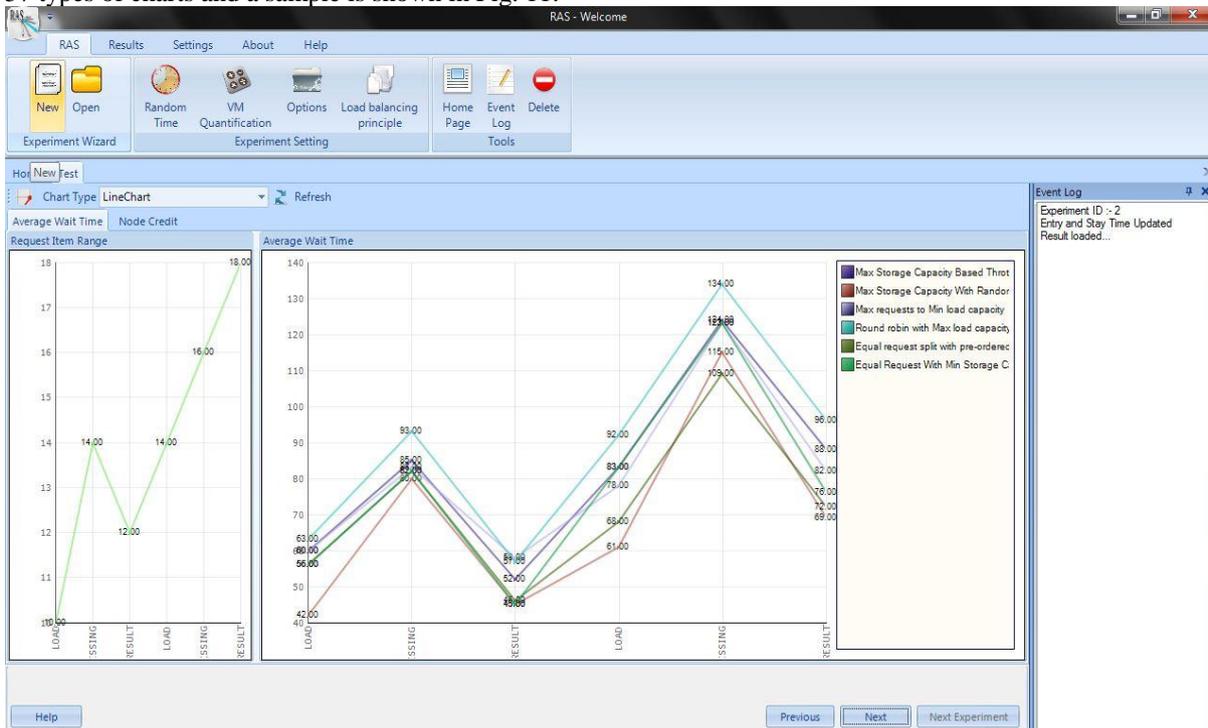

Figure 11. Comparison of performance

*F. Comparison of consolidated output parameters*

RAS consolidates each experiment's result with the already experiments under the same file name. When the number of experiments repeated with the same configuration set up will identify the strategy which yields better results. Fig. 12 shows the comparison of value earned by each node and average response time for all the requests till the current experiment. Therefore, the user can evaluate the consolidated output of all the experiments with a same set of configuration but varying inputs.

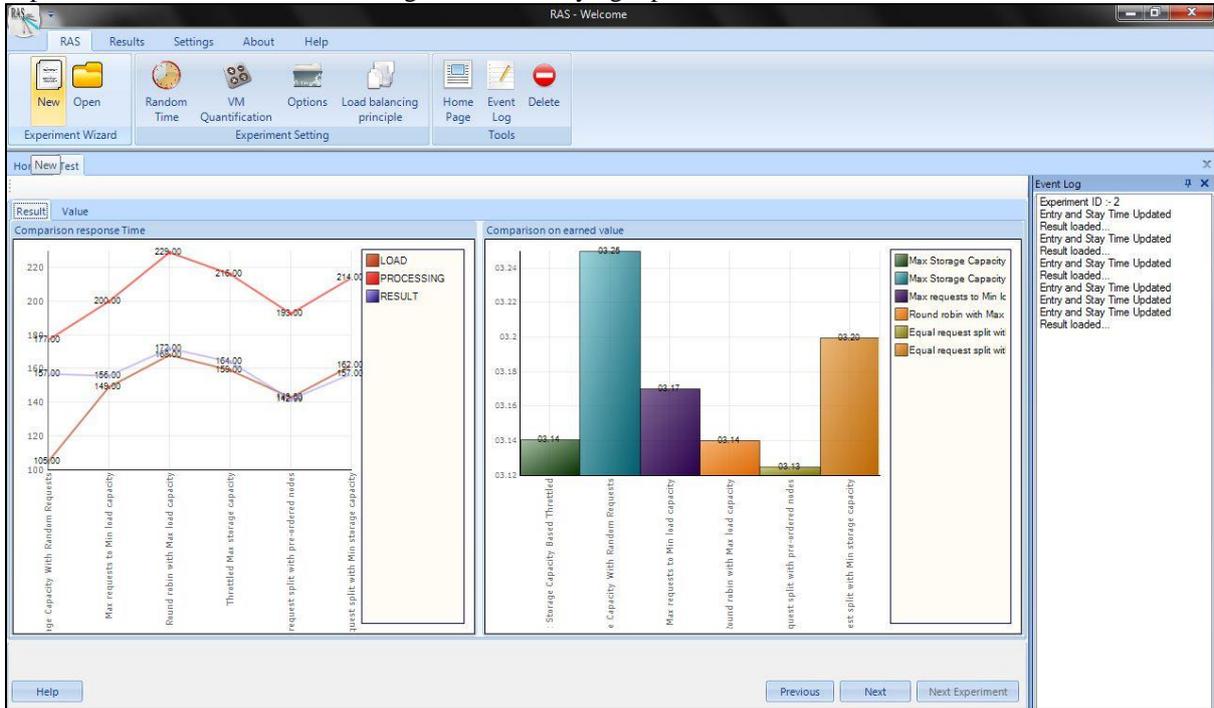

Figure 12. Comparison performance in a series of experiments

*G. Ordering of strategies based on the performance*

RAS computes resource utilization factor and average response time as the performance parameters. Resource utilization factor is the metric which assess the actual performance of the node with its expected performance. Fig. 13 presents the performance order of strategies based on the resource utilization factor. If the user wants, he can export the data as well as the charts into Excel.

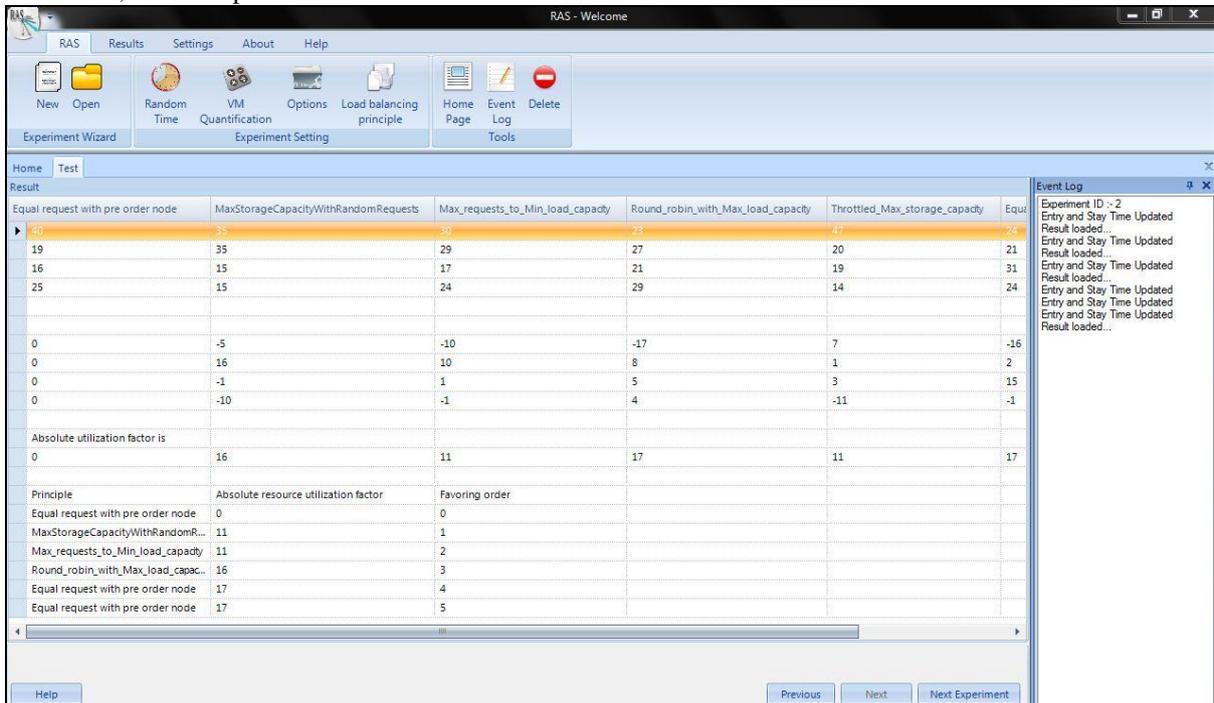

Figure 13. Ordering of strategies

With this an experiment is over. If the user wants to continue to experiment with the same configuration set up, he has to click "Next experiment". When it is clicked, the user has to input the number of inputs for all the services. The entered number of requests will be added to the previous demand value for the service. Accordingly the value for a service will be recomputed and automatically assigned in the cloud dashboard. Therefore the second experiment also has the same configuration set up, but the number of requests will be varying. The user is allowed to make changes to the configuration; otherwise RAS will assume the previous experiment set up and proceed further.

Before the conduct of the experiment, the user is advised to change his preferences in the configuration panel given as the tool boxes. The assembly contains different tools that are assisting the experiments.

*1) Random Time Settings*

Using the window of random time setting, the user can fix the range of time intervals for the arrival time and process time of requests. The generation of inputs is based on these range of values. Once the arrival time is set for an experiment with a range, then the next experiment's arrival time will start with the current experiment's upper bound value. The user only has to give the upper bound value of the arrival time for the new experiment. For example, if arrival time ranges from 7 to 18 for experiment ID 2, then experiment ID 3 will be automatically assigned by arrival time as 18 and users has to give the upper bound value.

When the user sets this range, it is suggestible to give in the range of 10 to 20. If the range is small, there won't be any big differences in the performance of the strategies.

*2) Strategy Assembly*

The Strategy Assembly, which lists the load balancing principles we experimented in RAS. Before starting an experiment, the user is supposed to select the principles of his interest. Fig. 14 shows the lists of load balancing principles from which the user can choose any principles for the current experiment. Currently RAS supports 27 load balancing principles with the following broader classification:
1. Round robin principle
2. Capacity fill-in principle
3. Throttled principle
4. Equal requests split
5. Capacity proportioned principle

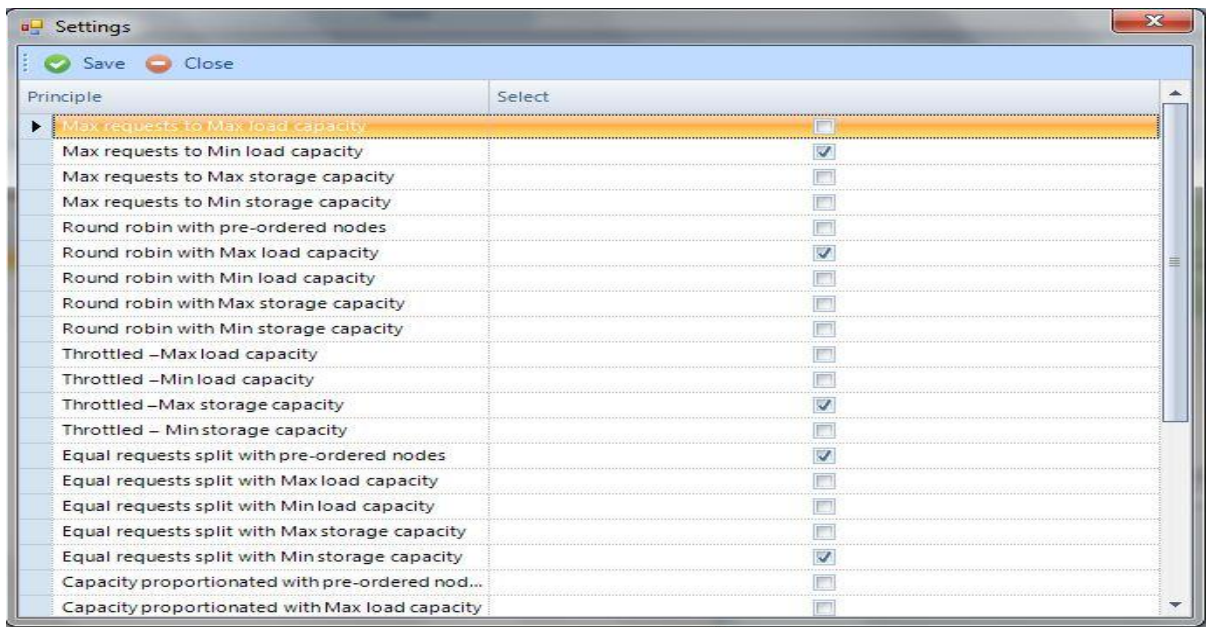

Figure 14. List of load balancing principles

*3) Options*

Using this window, the user can set various options to include specific conditions with the request assignment process. It includes options for priority setting, handling of the faulty virtual machine, and permitting the user to choose their processing zone.

*4) VM Quantification*

This component is for finding the percentage of requests to allocate with the virtual machines, according to the system's configuration. Each machine is quantified based on its load capacity or storage capacity. We used

Z-score statistical method for quantification process. Fig. 15 shows how to compute the quantification process for the current load capacity.

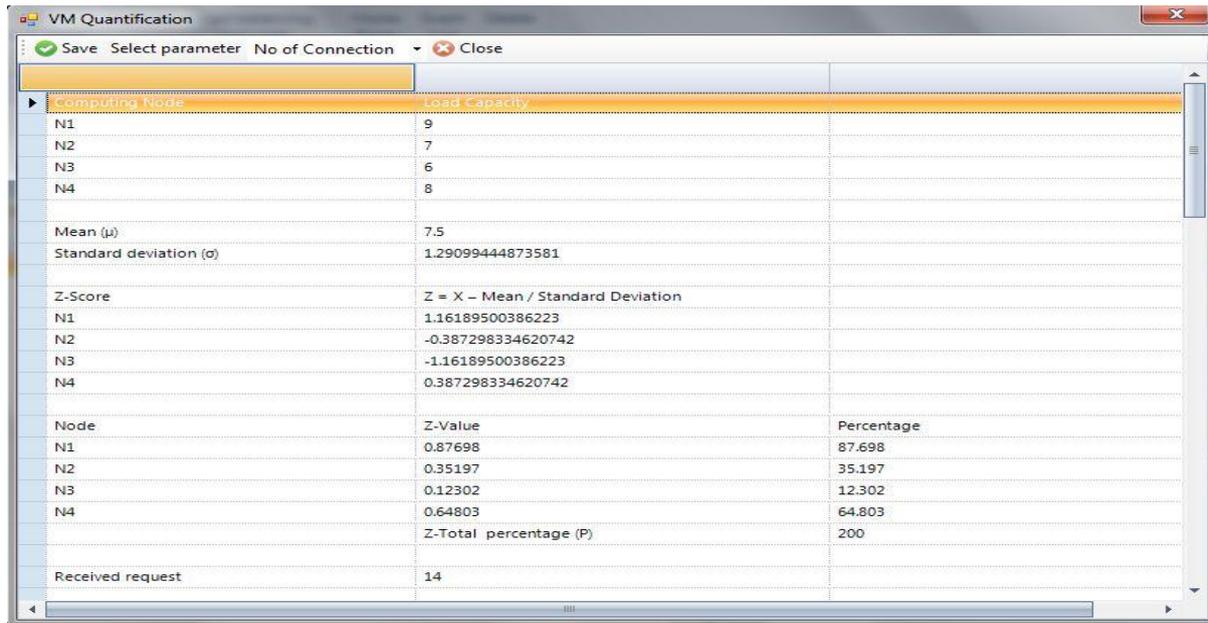

Figure 15. Virtual machines' quantification method

## IV. CONCLUSION

The objective of the load balancing principles is to assign the requests effectively with the computing resources. This is a challenging task in a large distributed computing environments like cloud. To the best of our knowledge, there is no existence of dedicated simulators to experiment cloud load balancing principles. As part of our research, we developed a novel simulator namely, Requests Assignment Simulator (RAS). In this paper, we presented a complete working manual for the user whoever wants to experiment with this tool. Because, even a best utility cannot serve its purpose if the user doesn't know the means of using it. We are confident that our simulator can be an experimenting tool for the academia and researchers.

We designed our simulator to accommodate any size of inputs. But it is suggestible to run the application with simple values. Once the user is comfortable with the working method of the simulator, he can try with large inputs. We designed our simulator based on the cloud simulator Cloud Analyst. It will be a remarkable progress if the parameters are further extendable to suit more relevantly with cloud architectures.

AUTHORS PROFILE

R Arokia Paul Rajan is working as Associate Professor at Pope John Paul II College of Education, Pondicherry as well as a research scholar in the Department of Computer Science & Engineering at Pondicherry Engineering College, Pondicherry, India.

F Sagayaraj Francis is working as Associate Professor in the Department of Computer Science & Engineering at Pondicherry Engineering College, Pondicherry, India. He holds Ph.D in Data Management from Pondicherry University, Pondicherry, India. He authored 15 journal and 13 conference publications. Some of his areas of interest include Database Management Systems, Data Mining and Knowledge Discovery, Data Structures and Algorithms, Knowledge and Intelligent Systems and Automata Theory and Applications.